\documentclass[reprint, amsmath,amssymb, aps,
]{revtex4-2}

\usepackage[utf8]{inputenc}
\usepackage{ragged2e}
\usepackage{array}
\usepackage{comment}
\usepackage{graphicx}
\usepackage{multirow}
\usepackage{amsmath}
\usepackage{amssymb}
\usepackage[none]{hyphenat}
\usepackage{xcolor}
\usepackage{hyperref}  
\usepackage{natbib}
\usepackage{tcolorbox}
\usepackage{enumitem}
\usepackage{tikz}
\usepackage{tikz-3dplot}
\usetikzlibrary{positioning}
\usepgflibrary{shapes.geometric}
\hypersetup{colorlinks=true,urlcolor=blue,citecolor=blue,linkcolor=blue}

\begin{document}
\sloppy
\preprint{APS/123-QED}

\title{Theoretical exploration of task features that facilitate student sensemaking in physics}

\author{Amogh Sirnoorkar}
\affiliation{Department of Physics, Kansas State University, Manhattan, KS 66506}

\author{James T. Laverty}
\affiliation{Department of Physics, Kansas State University, Manhattan, KS 66506}

\date{\today}

\begin{abstract}

Assessment tasks provide opportunities for students to make sense of novel contexts in light of their existing ideas. Consequently, investigations in physics education research have extensively developed and analyzed assessments that support students sensemaking of their surrounding world. In the current work, we complement contemporary efforts by theoretically exploring assessment task features that increase the likelihood of students sensemaking in physics. We identify the task features by first noting the salient characteristics of the sensemaking process as described in the science education literature. We then leverage existing theoretical ideas from cognitive psychology, education, and philosophy of science in unpacking the task features which elicit the characteristics of sensemaking. Furthermore, we leverage Conjecture Mapping -- a framework from design-based research -- to articulate how the proposed task features elicit the desired outcome of sensemaking. We argue that to promote sensemaking, tasks should cue students to unpack the underlying mechanism of a real-world phenomenon by coordinating multiple representations and by physically interpreting mathematical expressions. Major contributions of this work include: adopting an agent-based approach to explore task features; operationalizing conjecture mapping in the context of task design in physics; leveraging cross-disciplinary theoretical ideas to promote sensemaking in physics; and introducing a methodology extendable to unpack task features which can elicit other valued epistemic practices such as modeling and argumentation.

\end{abstract}

\maketitle

\section{Introduction}
\label{sec:intro}
% Start narrow by highlighting the cognitive outcomes of sensemaking.

Researchers in physics education have advocated for facilitating students' content understanding through promoting ``sophisticated epistemology'' - leveraging different modes of reasoning while engaging with a task~\cite{elby2001substance,bing2012epistemic}. The education research community has also emphasized promoting pedagogical practices that facilitate students in generating new knowledge by building on their existing ideas~\cite{brewe2008modeling}. Sensemaking~\cite{defining} -- the process of addressing a perceived gap in one's understanding -- attends to these valued objectives.

Sensemaking assists students' in better comprehending the curricular content by leveraging different forms of knowledge and practices~\cite{cannady2019scientific}. Sensemaking is also one of the many ways through which scientists and engineers generate new knowledge~\cite{ford2012dialogic,berland2009making,danielak2014marginalized}. Given this significance, there has been an uptick in investigations on the discourse markers and the nature of tasks associated with sensemaking. These include (but are not limited to) construction and critique of claims~\cite{ford2012dialogic}, vexing questions during interactions~\cite{odden2019vexing}, the blending of model- and evidence-based reasoning~\cite{russ2017intertwining}, computational reasoning about physics scenarios~\cite{computation}, addressing quantitative problems through qualitative insights and vice-versa~\cite{kuo2020assessing}, and explaining physical systems through mathematical insights~\cite{gifford2020curriculum}. We contribute to these efforts by theoretically exploring the assessment task features that increase the likelihood of students sensemaking in physics. 

We identify these features by initially noting  characteristics of the sensemaking process as described in the science education literature~\cite{defining}. Guided by the research in cognitive psychology, science education and philosophy of science, we make a theoretical argument for the task features that promote sensemaking in physics. We neither argue that the proposed features {\em necessarily} engage students in sensemaking nor any task that elicits sensemaking {\em necessarily} entails these features. We also do not claim that the proposed list is an {\em exhaustive} one. Rather, we make a modest argument that tasks entailing the proposed features {\em together} (as opposed to presence of one of these) increase {\em the likelihood} of students sensemaking. 

To highlight how the proposed features bring about the desired outcome of sensemaking, we elucidate the design criteria through a conjecture map. Conjecture mapping is a framework primarily employed in design-based research to conceptualize the interactions between theoretically salient design features of a learning environment and their intended outcomes~\cite{sandoval2014conjecture}. We adopt this framework to our context in elucidating how the proposed features elicit the desired outcomes of sensemaking. 

The current work makes four key contributions to the contemporary literature. Firstly, this study presents an agent-based approach in articulating the task features by shifting the vocabulary from ``{\em tasks entailing a feature X}''  to ``{\em tasks that cue students about X''} or ``{\em tasks that cue students to do X}''. Such vocabulary would better account for students' agency along with the local practices of their learning environments. Secondly, this work operationalizes the Conjecture Mapping framework in the context of task-design in physics. Thirdly, we leverage cross-disciplinary theoretical ideas particularly from cognitive psychology and philosophy of science in identifying the task features that promote sensemaking in physics. Lastly, our methodological approach in identifying task features can also be potentially extended in unpacking task features which can elicit other valued epistemic practices such as modeling and argumentation.

In doing so, we address the following research questions in the rest of this paper:

\begin{description}[noitemsep]
    \item[RQ1] How can we adopt a framework-based approach in theoretically identifying task features that promote students sensemaking in physics?
    
   \item[RQ2] What set of assessment task features increase the likelihood of students sensemaking in physics?
\end{description}

This manuscript is structured as follows: in the next section, we briefly review the literature on sensemaking before describing the theories of sensemaking and conjecture mapping in Sections~\ref{subsec:conjecture} and~\ref{subsec:sensemaking}. In Sections~\ref{sec:task-features}-\ref{sec:physical-interpretation}, we detail the arguments in support of the task features which increase the likelihood of students sensemaking in physics. We substantiate each argument by providing a theoretical background and empirical evidence from the literature. We conclude by discussing the implications and limitations of this work in Sections~\ref{sec:discussion} and~\ref{sec:conclusion}.

\section{Literature Review on sensemaking}
\label{sec:lit-review}

Science education literature has a rich repository of investigations on students sensemaking about their surrounding world. These explorations broadly span across three domains:  (i) theoretical descriptions of the sensemaking process (ii) analytical accounts exploring approaches of sensemaking, and (iii) the outcomes of sensemaking. We present a brief overview of the studies in each domain, and encourage readers to go through references~\cite{defining,zhao2021development} along with the cited literature for additional details.       

\subsection{Theoretical accounts of sensemaking}
\label{subsec:theoretical-sensemaking}

The first domain of the sensemaking literature has focused on theorizing the underlying process involved in `making sense' of a given context. These accounts have explored sensemaking through the lens of  transfer~\cite{nokes2013toward}, modeling~\cite{chen2022epistemic,russ2017intertwining,schwarz2009developing,schwarz2017helping,k-12framework,passmore2014models,sands2020modeling,gifford2020categorical}, argumentation~\cite{ford2012dialogic,ford2008disciplinary,ford2008grasp}, epistemic frames~\cite{zohrabi2020processes,hutchison2010attending}, and epistemic games~\cite{sensemakinggame}.   

According to Nokes-Malach and Mestre~\cite{nokes2013toward}, sensemaking forms a critical component of `transfer' -- the process of leveraging existing knowledge in solving novel problems. The authors argue sensemaking (during problem-solving) to be an iterative process involving coordination between prior knowledge and contextual information while generating an optimal solution. The process of narrowing down on the optimal solution is often achieved by `modeling' the given problem~\cite{chen2022epistemic,russ2017intertwining,schwarz2009developing,schwarz2017helping,k-12framework,passmore2014models}. Modeling as sensemaking entails an initial construction of mental models, and subsequent validation of the models' ideas through external representations~\cite{sands2020modeling}. One can also model the given context by employing mathematics as either a tool and/or as an object of investigation~\cite{gifford2020categorical}. Choosing the optimal solution candidate during sensemaking is also achieved through construction and critique of claims during an argument~\cite{ford2012dialogic,ford2008disciplinary,ford2008grasp}. Sensemaking from the argumentation perspective entails generation and evaluation of new knowledge, both at the individual, and at the community level.     

The idea of generating new knowledge is also resonated in other studies theorizing sensemaking as an `epistemic frame' (a tacit understanding of `what's going on here?'~\cite{hammer2005resources,sirnoorkar2016students}), or as an `epistemic game' (a strategic approach in perceiving an inquiry~\cite{Ferguson,tuminaro2007elements}). The sensemaking epistemic frame involves generation of novel explanations in response to a perceived gap in one's understanding about an observed phenomenon. These explanations are based on one's lived experiences, and often are aimed at unpacking the underlying mechanism that gives rise to the phenomenon~\cite{rosenberg2006multiple,hutchison2010attending,danielak2014marginalized,kapon2017unpacking}. On the other hand, the Sensemaking Epistemic Game~\cite{sensemakinggame} conceptualizes sensemaking as a multi-stage iterative process with a goal of addressing one's knowledge gap by leveraging contextual information and existing ideas.           

\subsection{Analytical accounts of sensemaking}

The second domain of the sensemaking literature has focused on analytically identifying reasoning approaches, or instances (mainly involving mathematics) that qualify as sensemaking~\cite{emigh2019equipotential,lenz2020sensemaking,kuo2020assessing,dreyfus2017mathematical,danielak2014marginalized,homework,hahn2019relativity,kuo2013students,gupta2011beyond,sherin2001students}. A review of this literature reveals a variety of definitions adopted to analyze the sensemaking process.  

A subset of this literature has defined sensemaking as establishing coherence between multiple representations of physics knowledge such as equations, figures, tables, or linguistic phrases~\cite{emigh2019equipotential,lenz2020sensemaking,danielak2014marginalized}. While Emigh {\em et al.}~\cite{emigh2019equipotential} define coordination between these forms of representations as sensemaking, Lenz {\em et al.}~\cite{lenz2020sensemaking} observe sensemaking as seeking coherence or meaning between them.  

Other studies have defined sensemaking as establishing connections between the structure of mathematical formalisms, and the physical world~\cite{kuo2020assessing,dreyfus2017mathematical,hahn2019relativity,homework,kuo2013students,gupta2011beyond,sherin2001students}. These studies have observed `mathematical sensemaking' to entail mapping of formal mathematics with causal relations~\cite{dreyfus2017mathematical}, conceptual understanding~\cite{kuo2020assessing,homework}, or intuitive reasoning~\cite{hahn2019relativity} about physical systems.

\subsection{Cognitive outcomes of sensemaking}
\label{subsec:outcomes}

The third domain of the sensemaking literature focuses on probing the cognitive outcomes of the sensemaking process. This literature posits three major outcomes of sensemaking: (i) generation of new knowledge, (ii) development of sophisticated epistemology, and (iii) enhanced content understanding.   

\subsubsection{Generation of new knowledge}

The first cognitive outcome of sensemaking is the generation of new knowledge by blending curricular ideas with lived experiences. Studies discussing episodes of sensemaking have noted students making novel claims by constructing analogies, making assumptions, designing thought experiments, and predicting outcomes~\cite{chin2000learning,ruibal2007physics,kapon2012reasoning,krist2019identifying}. Furthermore, sensemaking also entails a crucial component of scientists' and engineers' reasoning in knowledge construction while solving cross-disciplinary real-world problems~\cite{ruibal2007physics,nokes2013toward, danielak2014marginalized}. 
   
\begin{figure*}
     \centering
     \includegraphics[scale=0.475]{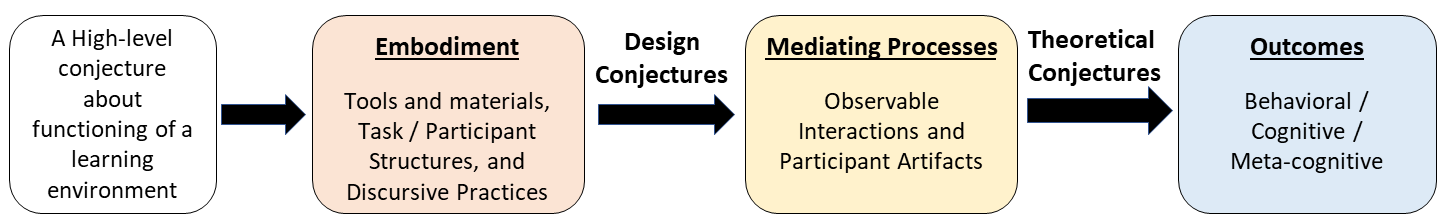}
     \caption{Modified schematic representation highlighting relationships between the elements of a conjecture map. The original representation can be found in~\cite{sandoval2014conjecture}. }
     \label{fig:conjecture-map}
\end{figure*}

\subsubsection{Sophisticated Epistemology}

Personal epistemology -- perspectives about what it means `to know', and the nature of knowledge -- plays a crucial role in how one engages with a given task~\cite{scherr2009student,elby2010epistemological,disessa1993toward}. During sensemaking, students iteratively coordinate and reconcile between different forms of knowledge and reasoning approaches. The knowledge forms include lived experiences, intuitive arguments, conceptual and procedural ideas, or hypotheses~\cite{defining,chin2000learning,danielak2014marginalized}. These knowledge forms are further accompanied with reasoning practices such as argumentation~\cite{ford2012dialogic}, problematization~\cite{odden2019vexing}, or modeling~\cite{sands2020modeling,sirnoorkar2022sensemaking}. This virtue of leveraging different forms of knowledge and blending them with a broad spectrum of epistemic practices results in sophisticated epistemology~\cite{bing2012epistemic,sirnoorkar2020qualitative} - the second major cognitive outcome of sensemaking.

\subsubsection{Enhanced content understanding}

One of the consequences of generating new knowledge through sophisticated epistemology is enhanced content understanding (the third cognitive outcome). Referring multiple sources of knowledge and leveraging varying epistemic practices during sensemaking contributes towards better content understanding~\cite{monte2011beyond,cervetti2014entering,fitzgerald2019teaching,sirnoorkar2016students,sirnoorkar2020towards,bing2009analyzing} by equipping students to `transfer' skills across multiple disciplines~\cite{cannady2019scientific,ruibal2007physics,nokes2013toward}.

\section{Theory}

\subsection{Agentic paradigm in task-design}

Research on task-design has traditionally involved prescribing a set of design features (often backed by analysis of students' responses) which can elicit a targeted response from students. However, studies have increasingly highlighted the role of contextual factors such as local norms of students' community (e.g., teachers, classrooms, and institutions) on what counts as ``knowing'' or ``doing'' science~\cite{stroupe2019introduction}, students' agency in accessing knowledge sources~\cite{miller2018addressing,vaughn2020student}, and their in-the-moment framing of the task's expectations~\cite{shar2020student} as influencing students' engagement with tasks. Efforts on explicit prompting in tasks too have evoked mixed results. While few have noted explicit prompting to enhance students' understanding on domain principles and procedural knowledge~\cite{berthold2011double,berthold2009instructional,nokes2011testing}, others have noted them to impede students' intuitive reasoning~\cite{heckler2010some} by selectively emphasizing parts of the presented information~\cite{grosse2006effects,rittle2017eliciting}. As Berland {\em et al.}~\cite{berland2016epistemologies} note

\begin{quote}
 ``{\em [...]emphasizing the actions alone can result in rote performance and attainment of skills, rather than student engagement in the rich work of scientific knowledge construction, evaluation, and refinement.''}
 \end{quote}

In light of these observations, we adopt an agent-based approach in arguing about task features by shifting the vocabulary from ``{\em tasks entailing a feature X}''  to ``{\em tasks that cue students about X}'' or ``{\em tasks that cue students to do X}''. By ``cuing'' we mean, conveying or setting up expectations for students about a feature in a task or about a specific way of reasoning as a solution approach to the task. Such vocabulary would better account for students' agency along with the local practices of their learning environments. In the rest of this paper, we adopt this framing in theorizing assessment features that can increase the likelihood of students sensemaking in physics.

\subsection{Conjecture Mapping}
\label{subsec:conjecture}

Design-based research accompanies a set of epistemic commitments about design and functioning of learning environments in addition to advancing the understanding of teaching and learning processes~\cite{sandoval2014conjecture}. Attending to these commitments often requires researchers to articulate conjectures about how the designed learning environment functions in an intended setting. Conjecture mapping~\cite{sandoval2014conjecture} is a technique which conceptualizes these arguments by establishing relationships between the design features, processes enacted by participants engaging with these features, and the intended outcomes. This technique highlights the relationships between various aspects of educational design through six elements: (i) a high-level conjecture, (ii) embodiment, (iii) mediating processes, (iv) outcomes, (v) design conjectures, and (v) theoretical conjectures (Figure~\ref{fig:conjecture-map}). 

A {\em high level conjecture} forms the first element of a conjecture map which articulates the theoretical idea driving the design of a novel learning environment. The articulated conjecture provides the road-map of the theoretical idea's operationalization in a given setting. This conjecture is then reified in {\em embodiment}, the second element of a conjecture map, which crystallizes the design features into several components. These components include: tools and materials (assessments, devices, etc.), task structures (the nature and form of tasks), participant structures (roles and responsibilities of participants), and discursive practices (forms of participants' discourses). These components further contribute to the {\em mediating processes}, a set of interactions and artifacts produced from the participants that mediate between the designed features and the intended cognitive/meta-cognitive {\em outcomes}. 

The {\em embodiment, mediating processes,} and {\em outcomes} are connected through {\em design} and {\em theoretical conjectures } - the last two elements of a conjecture map. Design conjectures are the arguments about how the components of embodiment (tools/materials, task/participant structures and discursive practices) lead to the mediating processes. Theoretical conjectures, on the other hand, are the arguments describing how the mediating processes will in turn result into the desired outcomes. Figure~\ref{fig:conjecture-map} schematically represents the elements  of a conjecture map and their interrelationships. 

We adopt conjecture mapping to elucidate how a set of task features (embodiment) can nudge students to engage ``sensemaking elements'' (mediating process) leading to generation of new knowledge, sophisticated epistemologies, and enhanced content understanding (outcomes). The theoretical arguments in favour of these outcomes (theoretical conjectures) are discussed in Section~\ref{subsec:outcomes}. Sections~\ref{sec:real-world} to~\ref{sec:physical-interpretation} detail the arguments about how the proposed task features elicit the features of sensemaking (design conjectures). Figure~\ref{fig:conjecture-map-task-features} represents the adoption of conjecture mapping to our study.

\subsection{Sensemaking}
\label{subsec:sensemaking}

Studies in science education have described sensemaking in diverse ways. In the rest of this paper, we adopt Odden and Russ'~\cite{defining} synthesized account of sensemaking as:

\begin{quote}
{\em a dynamic process of building or revising an explanation in order to `figure something out' - to ascertain the mechanism underlying a phenomenon in order to resolve a gap or inconsistency in one's understanding. One builds this explanation out of a mix of everyday knowledge and formal knowledge by iteratively proposing and connecting up different ideas on the subject. One also simultaneously checks that those connections and ideas are coherent, both with one another and with other ideas in one's knowledge system.}
\end{quote}

Odden and Russ put forward this definition by synthesizing three approaches through which researchers have conceptualized sensemaking in science education. In the first approach -- as a stance towards science learning -- sensemaking has been noted to entail generation of explanations describing the underlying mechanism of a phenomenon. In the second approach -- as a cognitive process -- sensemaking has been noted in involve integration of prior knowledge (experiences) with formal knowledge. In the last approach -- as a discourse practice --  sensemaking has been conceptualized as construction and critique of claims during argumentation. The construction component of argumentation entails proposing and connecting ideas to substantiate a claim. The critique component on the other hand, entails ensuring coherence between various the connected ideas.     

Based on the definition of the sensemaking process, and the conceptualizations of sensemaking across the three approaches in the science education literature, we note the following ``sensemaking elements'' or the set of activities crucial for engaging in sensemaking:

\begin{enumerate}
    \item Use of everyday and formal knowledge while reasoning about a phenomenon (sensemaking as a discourse practice). 
    
    \item Ascertaining the underlying mechanism of the phenomenon (sensemaking as a stance towards science learning).
    
    \item Generating and connecting up different ideas in one's knowledge system (sensemaking as a discourse practice).
    
    \item Seeking coherence between the generated ideas (sensemaking as a discourse practice).
\end{enumerate}

It should be noted that the above elements do not take into account the crucial aspect of noticing inconsistencies in one's understanding during sensemaking. The noticing of a discrepancy in one's knowledge system is a highly contextualized activity influenced by various factors including prior knowledge, awareness, self-evaluation, and adopted strategies while reasoning about a given scenario~\cite{anderson2022grasping}. Our list of sensemaking elements does not include this critical feature due to its highly contextual nature. In order to address this shortcoming in our theoretical approach, we adopt a probabilistic stance (``the task features {\em increase the likelihood} of students sensemaking'') rather than a deterministic one (``the task features {\em elicit} sensemaking'') in our arguments. 

We blend the above-mentioned sensemaking elements with conjecture mapping framework in identifying task features which promote sensemaking. By definition, if students engage in all of the above-mentioned sensemaking elements during an activity, they are more likely to engage in the sensemaking process. Along the same lines, we posit as our high level conjecture that {\em a set of task features which elicit the sensemaking elements increase the likelihood of students sensemaking}. Our design conjectures correspond to the arguments (articulated in Sections~\ref{sec:task-features} to~\ref{sec:physical-interpretation}) which link the proposed task features to the sensemaking elements of sensemaking. 

% \subsection{Self-explanations}

% \note{Generating explanations is a critical component of scientific reasoning and one of the eight scientific practices recommended to be promoted in science classrooms~\cite{k-12framework}. Generating explanations in response to new information or observations has been noted to be a strong predictor of students' learning and consequently been recommended as a study practice~\cite{dunlosky2013improving}.} 

% \note{A form of explanation generation -- self-explanations -- has attracted considerable traction in the education research community~\cite{chi1994eliciting}. Self-explanations are explanations to oneself generated while making sense of new information or observation. Self-explanations assist students in integrating different pieces of information with existing knowledge~\cite{lombrozo2006structure}. Self-explanations also present opportunities for students to notice potential conflicts between various pieces of information and resolve them~\cite{chi2013self}.      
% }

% \note{We adopt the theoretical idea of self-explanations in determining the scope and objective of task features that promote sensemaking. Since the identified sensemaking elements are intrinsically tied to explanation generation, particularly self-explanations, we consider that any task with an objective of eliciting sensemaking should also elicit self-explanations.}

\section{Task features that facilitate sensemaking}
\label{sec:task-features}

In Section~\ref{subsec:sensemaking}, we identified four ``sensemaking elements" or a set of activities which together contribute to the likelihood of sensemaking. These include: blending everyday and formal knowledge while reasoning about a phenomenon, ascertaining the underlying mechanism of the phenomenon, generating and connecting diverse ideas, and seeking coherence between the generated ideas. We posit that the set of task features which elicit these sensemaking elements increases the likelihood of students sensemaking in physics. 

%As noted in Section~\ref{sec:intro}, the literature on task-design has increasingly emphasized the role of contextual factors such as students' local norms on learning~\cite{stroupe2019introduction} their agency in accessing various forms of knowledge~\cite{miller2018addressing}, and their framing of the tasks' expectations~\cite{shar2020student} as influential in students' engagement with tasks. Consequently, we adopt an agent-based perspective in articulating task features by shifting the vocabulary of discourse on task features from ``{\em tasks entailing a feature X}''  to ``{\em tasks that cue students about X or to do X}''. This change in stance takes into account not only the valued task features by the education research community but also on how students perceive the task based on their contextual factors.   

In the next four sections, we propose that tasks which cue students about the following to promote sensemaking in physics: (i) the presence of real-world context(s), (ii) to engage in mechanistic reasoning, (iii) to coordinate between multiple representations, and (iv) to extract physical implications from mathematical expressions. Each section consists of {\em conjectures} - arguments about a task feature eliciting specific sensemaking elements, {\em theoretical background} - a theoretical basis of the argument, and {\em empirical evidence} - evidence in favor of the argument from the literature. Table~\ref{tab:design-conjectures} summarizes these components.  

\section{Tasks cuing about the presence of real-world context(s)}
\label{sec:real-world}

\begin{figure*}
     \centering
     \includegraphics[scale=0.52
     ]{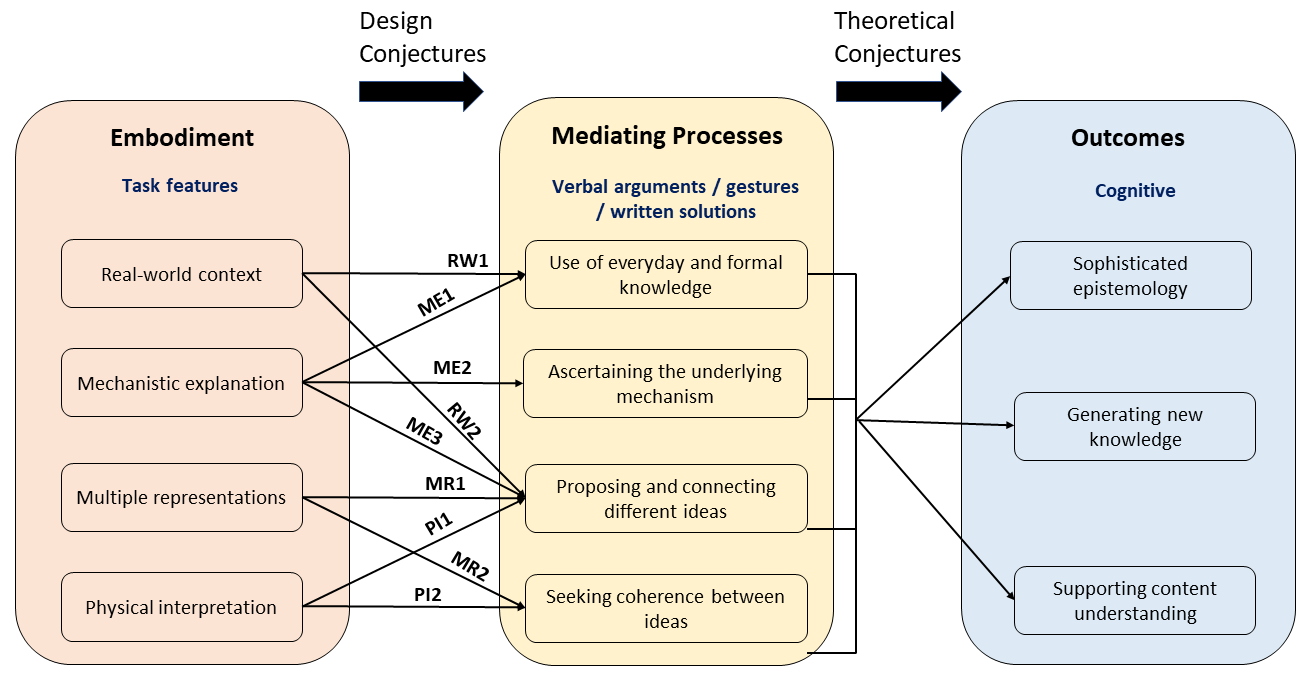}
     \caption{Contextual operationalization of the conjecture map in our study. Our high level conjecture (not represented in this figure) takes the form: ``a set of task features which elicit the sensemaking elements increase the likelihood of students sensemaking". While the design conjectures are detailed in Sections~\ref{sec:real-world}-~\ref{sec:physical-interpretation}, the theoretical conjectures are discussed in Section~\ref{subsec:outcomes}.} 
     \label{fig:conjecture-map-task-features}
\end{figure*}

The first feature we argue to contribute for students sensemaking is the task cuing students about the presence of real-world context(s). In line with the contemporary discourse in the science education literature, we consider a real-world context as a scenario relevant to the learner, and which requires application of scientific principles/models to make sense of the presented scenario~\cite{loffler2018students}. We argue that tasks perceived as rooted in real-world contexts facilitate two of the four sensemaking elements of sensemaking: use of everyday and formal knowledge; and generating and connecting up diverse ideas in students' knowledge system. These arguments, as design conjectures in our conjecture map (Figure~\ref{fig:conjecture-map-task-features}), have been labeled as RW1 and RW2.      

\subsection*{RW1: Real-world contexts facilitate use of everyday and curricular knowledge}

{\bf Conjecture RW1:} {\em If a task cues students about the presence of a real-world context, then it is more likely to invoke their everyday and curricular knowledge.} In other words, we posit that real-world scenarios in physics tasks  appeal to students' lived experiences along with priming their formal curricular ideas. In order to substantiate this argument, we turn to studies in cognitive psychology probing the influence of words or phrases in tasks priming specific information from one's knowledge system. 

{\bf Theoretical background:} Investigations on human interactions with tasks associated with a language's vocabulary (lexical tasks) have observed the role of  tasks' contexts on participants' reasoning~\cite{eich1985context}. According to these studies, the greater the relevance of the task's context to the participants, the better is the task's interaction with their memories~\cite{smith1978environmental,godden1975context}. `Semantic priming'~\cite{hutchison2013semantic} is one of the theoretical constructs proposed to explain how words or phrases in a lexical task cue related ideas from one's memory.

Semantic priming is a cognitive effect in which people respond faster to targeted words (e.g., `dolphin') when they are preceded by related words (e.g., `whale'), as compared to the unrelated ones (e.g., `chair'). Semantic relatedness represents the similarity in meaning or the overlap in featural description between a set of words or phrases~\cite{ferrand2004semantic}. Collins and Loftus~\cite{collins1975spreading}, present the `Spreading Activation Theory' to describe the mechanism through which semantic memory is accessed during lexical activities. According to this theory, semantic memory consists of a network of interconnected nodes, with each node representing a concept. A ``concept'' can take several forms ranging from a word to a proposition. The connections between any pair of nodes represent the information connecting the two concepts. The stronger the connections between the two nodes, the easier it is to retrieve associated concepts from memory.

This memory network further embeds a network called semantic networks where the nodes are connected based on the words' meaning, and their shared features.  The strength of association between the nodes depends on the degree to which the associated nodes share common features. For instance, the semantic association of the word ``red'' is stronger with ``rose'' as compared with ``elephant''. When a concept is primed during a lexical activity (such as while reading the task prompt), activation spreads out from the primed node along the paths of the network. The ``intensity'' of activation spread is higher for a strongly associated pair of nodes.

We conjecture that context-based tasks trigger semantic priming with activation spread emanating from concepts (nodes) associated with students' lived experiences as well as with their curricular knowledge. In other words, real-world scenarios in tasks are more likely to invoke arguments from everyday lives and formal knowledge. This cuing is more likely to be semantic in nature, i.e., based on shared features of the words/phrases in the task description. 

{\bf Empirical evidence:} We find empirical evidence for our above conjecture from several studies in PER. Odden and Russ~\cite{odden2019vexing}, while noting the role of vexing questions in sensemaking, discuss a pair of students' (Jake and Liam) reasoning on a task rooted in real-life. The task inquires about the safety of a car's passengers when exiting the vehicle following a lightning strike during a thunderstorm (with the passengers inside the car being unaffected by the lightning). The students approach the task by blending conceptual arguments about charge distribution with their everyday experiences about the shape of the car door's handle. Similar observations reflecting amalgamation of curricular knowledge with lived experiences can be found in case studies involving context-based tasks from other studies in PER~\cite{methodssirnoorkar,sirnoorkar2022sensemaking}. 

A more direct evidence for our conjecture comes from Enghag {\em et al.}'s~\cite{enghag2007everyday} study exploring students' reasoning about a context-rich physics problem. The authors note students initiating their approaches by rephrasing the given prompt based on their lived experiences before referring to the underlying physics principles. The authors highlight references to everyday knowledge as instrumental in students' meaning making, and understanding of the physics involved in the task.   

\subsection*{RW2: Real-world contexts facilitate generating and connecting diverse ideas}

{\bf Conjecture RW2:} {\em If a task cues students about the presence of a real-world context, then it is more likely to lead students to generate and connect diverse sets of ideas (conceptual, procedural and intuitive) from their knowledge system}. We again refer to the literature from cognitive psychology, particularly on search and selective retrieval of ideas from memory~\cite{madore2015creativity,madore2016divergent,storm2014forgetting} in  support of our argument. 

{\bf Theoretical background:} Nijstad {\em et al.}~\cite{nijstad2002cognitive}, propose the Search for Ideas in Associative Memory (SIAM) as a mechanism to describe how ideas get generated while engaging in an activity. According to this account, the internal process of idea generation proceeds through two distinct stages: (i) knowledge activation, and (ii) idea production. In the knowledge activation stage, a search in one's memory networks is triggered by a cue from the contextual features of the task. The structure and function of these memory networks is similar to the networks discussed in Activation Spread Theory in RW1. The memory search initiated by the contextual cue results in the retrieval of an image (idea), whose probability of retrieval depends on the strength of association between the cue and the image. In the second stage, i.e., the idea production stage, the initial image (produced in the previous stage) now acts as the triggering cue, leading to the production of an additional image. This chain of image production -- a preceding image acting as a triggering cue for a new image -- results into a ``train of thought'' until the information processing session is terminated. The conditions of termination depend on the nature and outcomes of the activity.   

We conjecture that presence of real-world contexts in physics tasks are more likely to trigger a diverse set of ideas from one's knowledge system. From the viewpoint of SIAM, this conjecture can be rephrased as: context-based tasks are more likely to trigger knowledge activation leading to generation of diverse `trains of thoughts'. This argument, as a design conjecture in our conjecture map (Figure~\ref{fig:conjecture-map-task-features}) has been labeled as RW2. 

{\bf Empirical evidence:}  We find several references in cognitive psychology and science education literature in support of our above argument. While discussing the SIAM account, Nijstad {\em et al}~\cite{nijstad2002cognitive} further note that semantically diverse cues (having diverse featural association between the cues) in a task lead to the generation of diverse set of ideas. George and Wiley~\cite{george2020need} note people ``rely too heavily on familiar or easily accessible information during idea generation''. Other researchers too have made similar observations on the familiarity of contextual cues stimulating generation of novel ideas~\cite{connolly1993effectiveness,kohn2011collaborative,fink2010enhancing}. In science education, Rennie and Parker~\cite{rennie1996placing} document students' perspectives on solving physics problems based on real-life scenarios. The authors note students referring to context-rich problems as ``easier to visualize'' as one of the emerging themes in students' responses. 

\section{Tasks cuing students to generate mechanistic explanation}

Mechanistic explanations -- descriptions unpacking the underlying mechanism of a phenomenon -- have been considered more sophisticated as compared to say, occult or teleological accounts~\cite{russ2008recognizing,kapon2017unpacking}. In what follows, we argue that tasks cuing students to generate a mechanistic explanation of a (real-wold) phenomenon to elicit three of the four sensemaking elements. These include: referring everyday and curricular knowledge, ascertaining the underlying mechanism of a phenomenon, and proposing and connecting up different ideas in one's knowledge system. These conjectures have been respectively labelled as ME1, ME2, and ME3 in Figure~\ref{fig:conjecture-map-task-features}. As the ME2 conjecture -- tasks requiring students to generate a mechanistic account lead to mechanistic reasoning -- is self explanatory, we will exclude it from detailed discussions below.  

\subsection*{ME1: Mechanistic explanations facilitate use of everyday and curricular knowledge}

{\bf Conjecture ME1:} {\em If a task cues students to generate mechanistic explanation(s), then it is more likely to invoke references to everyday and curricular knowledge.}  We substantiate our argument by referring to studies on storage and accessibility of knowledge about mechanisms in memory. 

{\bf Theoretical background:} The cognitive science literature argues for six possible formats through which knowledge about mechanisms (henceforth referred to as `mechanism knowledge') is internally represented. These include: (i) associations, (ii) forces or powers, (iii) icons, (iv) placeholders, (v) networks, and (vi) schemas. A detailed discussion about each of the representational formats would be beyond the scope of this paper. However, we briefly describe each of these formats, and encourage readers to go through ~\cite{glennan2017routledge,waldmann2017oxford,johnson2017causal} along with the cited references for additional details.

``Associations'' represent the mapping between two or more distinct events from one's memory such that the knowledge about a familiar event guides the expectations about the unfamiliar one~\cite{shanks1988associative,le2017associative}. In physics, this association can be observed in the ways propagation of sound in a medium is explained in terms of the compressions and rarefactions occurring on a vibrating spring. The second format, ``forces''~\cite{talmy1988force,wolff2007representing} or ``powers''~\cite{harre1975causal,white1989theory}, posits that mechanistic inferences are driven by the knowledge of physical laws. According to the ``forces'' account, interaction between entities (e.g., collision between two objects) are mediated by forces, and this interaction is described through vectors highlighting the direction of the entities' motion in the influence of the involved forces. On the other hand, the ``powers'' account posits that humans comprehend mechanisms by conceptualizing entities as having inherent dispositional features. These features either take the form of ``powers'' (tendency to bring about effects) or ``liabilities'' (tendency to undergo the effects). Melting of ice in presence of heat, for instance, would be explained in terms of the ``power'' of heat (causing the ice to melt) and the ``liability'' of ice (to melt in the influence of heat). 

The third candidate -- ``icons'' -- is a representation in which mechanisms are conceptualized as mental simulations or mental models consisting of a series of icons or image-like formats~\cite{barsalou1999perceptual,goldvarg2001naive,johnson2017mental}. The mechanistic imagery that humans possess about the functioning of gears or pulleys is an example of this format~\cite{schwartz1996shuttling}. On the contrary, the ``placeholders'' account (the fourth representational format) posits that people tend to hold a placeholder or a reference pointer for mechanisms instead of a detailed knowledge~\cite{rozenblit2002misunderstood}. Studies arguing for this format have observed people to possess skeletal details about the functioning of familiar everyday complex systems (such as sewing machines or can openers) with a meta-representational placeholder representing an unknown existing mechanism.

The penultimate representational format, ``networks'', has its origin in statistics and artificial intelligence. According to this account, causal relations are internally comprehended through causal networks (or ``Causal Bayesian Networks'') in which the nodes represent the variables involved in a mechanism, and the links between the nodes represent the causal relations between the involved variables~\cite{griffiths2009theory,pearl2000models,rottman2017acquisition}. As an example, the experience of drinking coffee leading to the sense of feeling energized would be represented in a typical causal network with ``drinking coffee'' and ``feeling energized'' as two nodes with an arrow pointing from the former towards the latter. The last candidate in our list -- ``schemas''~\cite{schank2013scripts} -- correspond to clusters of knowledge in the long-term memory that are employed while figuring out the mechanism of a phenomenon. For instance decisions on the appropriate container to carry cold drinks during summer, are guided by the schemas about  heat conductivity through various kinds of materials encountered in daily lives.

One of the common themes across the six representational formats discussed above is their association with one's prior knowledge. The formats highlight that people construct mechanistic accounts by building on their existing notions about the functioning of their surrounding world. Consequently, we posit that tasks cuing students to generate a mechanistic explanation, particularly about a real-world context/phenomenon, can nudge them to invoke their everyday ideas in addition to knowledge gained from formal instruction.  

{\bf Empirical evidence:} Several studies in physics education provide empirical evidence in support of our argument. For instance, diSessa~\cite{disessa1993toward} observes students to have a ``{\em sense of mechanism}'' through which they gauge the likelihood of various events, make ``backward and forward chaining'' of events~\cite{russ2008recognizing}, and provide the causal account of an observed phenomenon. This sense of mechanism is built from basic sensemaking elements called ``phenomenological primitives'' which are in turn derived from one's lived experiences. Resonating a similar view, Hammer~\cite{hammer1995student} notes students and physicists to have ``rich stores of causal intuitions'', and generating mechanistic explanations to entail references to lived experiences and formal ideas. Sirnoorkar {\em et al.}~\cite{sirnoorkar2022sensemaking,methodssirnoorkar} too observe student-generated mechanistic account of an amusement park ride (a real-world context) to involve an amalgamation of lived experiences and curricular ideas.

\subsection*{ME3: Mechanistic explanations facilitate generation and connection of different ideas}

{\bf Conjecture ME3:} {\em If a task cues students to generate a mechanistic explanation, then it is more likely to lead them in generation and connection of diverse ideas from their knowledge system}. We support our argument by discussing the nature and features of mechanistic reasoning as described in the philosophy of science and science education literature. To begin with, as noted above, mechanistic reasoning entails drawing ideas from lived experiences and curricular knowledge. Thus, intuitive and formal insights contribute to the spectrum of ideas invoked in unpacking the mechanism of a phenomenon. 

{\bf Theoretical background:} Furthermore, mechanistic reasoning is a complex cognitive process involving description of the behaviour of relevant entities and processes that give rise to a phenomenon~\cite{russ2008recognizing,krist2019identifying,louca2011quest,machamer2000thinking}. One generates mechanistic accounts by transitioning from observable features of the phenomenon at the macro level to the underlying entities or processes (often at the micro level)~\cite{russ2008recognizing,krist2019identifying}. The process of ascertaining the mechanism can further involve transitioning back from the micro to the macro features, and testing the validity of the generated explanations by varying the spatial or temporal organization of the entities or processes. This cyclic navigation across ``scalar levels" -- between observable features and underlying entities or processes, requires one to invoke conceptual, procedural or intuitive ideas and establish coherence between them. This argument as our design conjecture, has been labelled 'ME3' in Figure~\ref{fig:conjecture-map-task-features}.

{\bf Empirical evidence} Several studies describing episodes of mechanistic reasoning have noted students invoking and connecting diverse sets of ideas in their explanations~\cite{russ2008recognizing,de2022students,bachtiar2021stimulating}. Russ {\em et al.}~\cite{russ2008recognizing} discuss first-grade students' mechanistic account of a scenario involving a piece of paper and a book simultaneously dropped from a same height. The students explain the mechanism of falling objects in terms of gravity (a curricular idea) and everyday experience of jumping and landing back on the ground. Similarly, de Andrade {\em et al.}~\cite{de2022students} discuss a pair of middle school students' collaborative exploration of how antacid pills neutralize stomach's acidity. The students (Iris and Raul) generate an explanation by invoking the conceptual argument of the formation of salt and water upon the acid-base reaction. This argument also accompanies a procedural idea of the combination of elements during reaction in determining the molecular formula of the salt and water. The students also reason by making arguments based on everyday experiences that molecules (or objects in general) get smaller in size after collision in a reaction. We find a similar observation in Bachtiar {\em et al's.}~\cite{bachtiar2021stimulating} study in which students invoke conceptual, procedural and intuitive ideas while generating mechanistic accounts of a soccer ball's motion while designing its animation.         

\section{Tasks cuing students to engage with multiple representations}

Elucidating complex ideas through multiple external representations such as equations (wave functions, equations of state), graphs (kinematic plots, isotherms), or words (laws, theorems) is a common practice in physics. By multiple representations, we mean a combination of distinct external representations that illustrate the same content but use different symbol systems~\cite{rau2020cognitive,van2020handbook}. Representational formats of an idea complement each other by highlighting specific information about its content~\cite{gire2015structural,susac2017graphical,ainsworth1999functions,cox1999representation,ainsworth2006deft}. For instance, the kinematic equation $v= v_0 + at$ can better highlight the dependence of an object's final velocity ($v$) on initial velocity ($v_0$), duration of its motion ($t$), and its uniform acceleration ($a$). On the other hand, the graphical representation of the same equation (velocity vs time plot) better highlights the qualitative variation of the object's velocity for a given nature of acceleration (positive, negative or zero). 

We argue that tasks cuing students to engage with multiple representations -- either provided or constructed -- address the following sensemaking elements: proposing and connecting up different ideas; along with establishing coherence between them. These arguments as our design conjectures, are labelled ``MR1'' and ``MR2'' in Figure~\ref{fig:conjecture-map-task-features}.  As a primer, note that unlike the last two sections, in the current section and in the next one, we substantiate the the relevant design conjectures through a common theoretical background. 

{\bf Theoretical background:} As a basis for these conjectures, we refer to Mayer's ``Cognitive Theory of Multimedia Learning (CTML)''~\cite{mayer2005cognitive} describing the cognitive process involved in interacting with multiple representations. According to CTML, engaging with multiple representations (or multimedia) involves participation of sensory, working, and long term memories. Sensory memory is a short-term memory in which information obtained through sensory inputs (such as visuals of a painting) are stored in their original perceptual form. Working memory corresponds to the cognitive faculty involved in processing and manipulating instantaneous information in active consciousness (e.g., the cognitive process invested in comprehending the meaning of this sentence). Lastly, the long term memory corresponds to the accessible information stored across longer periods of time (e.g., information about one's childhood). 

With the participation of these memory forms, the cognitive process involved in interacting with multiple representations proceeds through three distinct and consecutive phases:  (i) {\em selection}, (ii) {\em organization,} and (iii) {\em integration} of information. As noted earlier, each representational format of an idea highlights a specific component of the information about the idea. The first phase -- {\em selection} -- involves selective choice of this information to be expressed into, or extracted from each representational format with the participation of one's sensory memory. In our kinematic example, the selective extraction of the information about the interdependence of the variables ($v$, $v_0$, $a$, and $t$), along with their behavior in limiting conditions, mark the {\em selection} phase associated with the algebraic representation. Similarly, an analogous argument can be made about the graphical representation ($v-t$ plot), in which the qualitative information about the velocity variation is selectively comprehended.  

The next phase -- {\em organization} -- involves forming mental representations of the embedded, or the selected pieces of information in the working memory. These mental representations are constructed by establishing internal connections between the informational pieces. In the kinematic example, this can correspond to the formation of mental representations of the interdependence of the variables (extracted from the equation), and the velocity variations for a given  acceleration (extracted from the graph). Lastly, these mental representations are fused with the help of prior-knowledge drawn from the long-term memory marking the {\em integration} phase of the CTML. In the kinematic analogy, this phase can correspond to the amalgamation of the algebraic and graphical mental representations using existing knowledge about slopes, or about uniform/non-uniform motion of objects. 

\subsection*{MR1: Engaging with multiple representations facilitate generation and connection of ideas}

{\bf Conjecture MR1:} {\em If a task cues students to engage with multiple representations, then it is more likely to lead students into generation and connection of ideas from their knowledge system.} Based on the CTML's three phases, particularly the {\em selection} and the {\em organization} phases, we note that engaging with multiple representations involve generation and connection of ideas. While the former phase involves idea generation through selective interaction with information from the representations, the latter involves connecting the ideas through formation of mental representations. 

{\bf Empirical evidence:} Several studies in physics education have made observations about representations facilitating generation and connection of ideas. Researchers have observed multiple representational formats to cue students in employing and connecting diverse set of domain-specific principles and strategies during problem solving~\cite{de2012representation,podolefsky2006use,van2001multiple}. De Cock~\cite{de2012representation} observes that an isomorphic task presented in varying representational formats tends to elicit different solution approaches along with physics principles. On a similar study, Podolefsky and Finkelstein~\cite{podolefsky2006use} note that use of multiple representations can facilitate mapping of ideas during analogical reasoning. Van Heuvelen and Zou~\cite{van2001multiple} note multiple representations of work-energy processes such as verbal descriptions, bar-charts, and mathematical equations facilitate students in better visualizing the energy conservation principle in addition to production of ``mental images for different energy quantities''. 

\subsection*{MR2: Engaging with multiple representations facilitate establishing coherence between ideas}

{\bf Conjecture MR2:} {\em If a task cues students to engage with multiple representations, then it is more likely to nudge them in seeking coherence between ideas}. Along the same lines, the CTML's last two phases -- {\em organization} and {\em integration} -- highlights that engaging with multiple representations facilitates establishing coherence between the generated ideas. While the former phase entails establishing coherence between the selected pieces of information from a representational format, the latter involves establishing coherence between ideas from representations. Seufert~\cite{seufert2003supporting} refers to these two phases as `intra-representational coherence formation' (establishing interrelations {\em within} a representational format), and `inter-representational coherence formation' (establishing interrelations {\em between}  representational formats).

{\bf Empirical evidence:} Cox~\cite{cox1999representation} argues that external representations help in better comprehending an idea as each representational format directs attention to a particular characteristic feature highlighted by the representation. Indeed, Gire and Price~\cite{gire2015structural} note students reasoning in quantum mechanics by effectively coordinating between Dirac, algebraic and matrix notations while representing quantum states of a system. The authors observe students establishing coherence between their ideas by using one notation as a template while creating corresponding representations in other notations.

\section{Tasks cuing students to extract physical implications from mathematical expressions}
\label{sec:physical-interpretation}

Physics education research has an extensive corpus of discussions on the role and use of mathematics in physics~\cite{hsu2004resource,docktor2014synthesis}. A major section of this work has analyzed students' interaction with mathematical formalisms during problem solving~\cite{sirnoorkar2016students,dreyfus2017mathematical,sirnoorkar2020towards}. In the rest of this subsection, we argue that tasks cuing students to extract physical implications from mathematical expressions (equations, plots, etc.) lead to generation and connection of ideas, along with establishing coherence between them. These arguments have been labelled ``PI1'' and ``PI2'' in our conjecture map (Figure~\ref{fig:conjecture-map-task-features}).   

{\bf Theoretical background:} Discussions in the philosophy of science literature posit that extracting physical implications from mathematical expressions involve mapping structural features of mathematical formalisms to that of physical systems~\cite{pincock2004revealing,ddi,bueno2011inferential,dfdif}. This view has been identified with several theoretical perspectives such as `mapping account'~\cite{pincock2004revealing}, `interpretation'~\cite{ddi}, `inferential conception'~\cite{bueno2011inferential} or `inferential function'~\cite{dfdif}. Nevertheless, the underlying theoretical view remains that interpreting mathematical relations involve bridging the structure of mathematical formalisms with the features of the target system. For instance, inferring the motion of a spring (target system) from the equation $F = -kx$ (mathematical structure) involves mapping the algebraic symbol $F$ with the net force on the spring, $x$ with the spring's displacement from its mean position, $k$ with the spring constant and the negative sign with the force's direction. Evidently, this mapping requires one to simultaneously engage with formal mathematical ideas (procedural or conceptual) along with ideas about the physical system. Consequently, we argue that the process of interpreting meaning from mathematical expressions involve generation of ideas and establishing coherence between them.  

\subsection*{PI1: Physical interpretations facilitate generation and connection of ideas}

{\bf Conjecture PI1:} {\em If a task cues students to interpret mathematical expressions in light of physical systems, then it is likely to facilitate generation and connection of ideas.}  Physically interpreting mathematical expressions is a common practice in physics. Whether it's determining the likelihood of an event based on the changes in entropy of involved systems, or identifying the position of an image from ray diagrams, students and physicists alike are familiar with this practice. 

{\bf Empirical evidence:} Several studies have noted interpretation of mathematical results as a crucial component of reasoning in physics~\cite{ACER,bing2009analyzing,tuminaro2007elements,redish2008looking}. Sherin~\cite{sherin2001students} makes a case for the existence of knowledge structures called `symbolic forms' which mediate the process of meaning making through mathematical formalisms. According to this view, students blend contextual ideas with mathematical insights while interpreting (or expressing) meaning from mathematical expressions. Making a similar observation, Arcavi~\cite{arcavi1994symbol} argues for `symbol sense' in mathematics, which facilitates interpretation of mathematical expressions via intuitions. 

Perhaps, a more direct evidence in support of our argument comes from the study by Kuo {\em et al.}~\cite{kuo2013students} investigating students' blending of conceptual arguments with formal mathematics. The authors discuss one of their participants' (Pat) reasoning  about the difference between final velocities of two balls dropped with differing initial velocities. The reasoning approach involves interpreting a kinematic equation ($v = v_0 + at$) through the lens of derivatives (a mathematical idea), and linking it to the variation of the balls' parameters (ideas of the physical system). Gifford and Finkelstein~\cite{gifford2020categorical} term this approach as mathematical sensemaking involving use of mathematical `tools' to reason about physical system.    

\subsection*{PI2 Physical interpretations facilitate establishing coherence between ideas}

{\bf Conjecture PI2:} e{\em If a task cues students to interpret mathematical expressions in light of physical systems, then it is likely to nudge them in seeking coherence between ideas.} Along the same lines, we argue that the process of interpreting mathematical expressions involves establishing coherence between the generated ideas. In our above example involving the spring's motion, one can interpret the negative sign in the equation as the net force and the displacement vectors being oppositely directed at a given instant of time. 

{\bf Empirical evidence:} Several studies in PER have indeed referred to the process of coherence seeking between mathematics and physical systems as ``mathematical sensemaking''. While Kuo {\em et al.} define it as ``{\em leveraging coherence between formal mathematics and conceptual understanding}''~\cite{kuo2020assessing}, Dreyfus {\em et al.} define the same as ``{\em looking for coherence between the structure of the mathematical formalism and causal or functional relations in the world}''~\cite{dreyfus2017mathematical}. Wilcox {\em et al.}~\cite{ACER} further note this practice as `Reflection of results' while discussing upper-division students' use of mathematics in physics.   

\renewcommand{\arraystretch}{1.3}
\begin{table*}
\begin{center}
\caption{A brief summary of the task features, design conjectures associated with each feature, theoretical background for the design conjecture and the corresponding empirical evidence from the literature.}
\label{tab:design-conjectures}

\begin{ruledtabular}

\begin{tabular}{p{2.5 cm} p{7.5 cm} p{0.2cm} p{3.5 cm} p{0.3cm} p{7 cm}}

Task-feature  & Conjecture && Theoretical basis  & & Empirical evidence \\
\hline

& If a task cues students  &&&& \\

\multirow{2}{2.5cm}{Real-world context (RW)} & (RW1) about the presence of a real-world context, then it is more likely to invoke students' everyday and curricular knowledge.  && Semantic priming and Spreading activation theory & &\cite{odden2019vexing,sirnoorkar2022sensemaking,enghag2007everyday} \\

 & (RW2) about the presence of a real-world context, then it is more likely to lead students to generate and connect diverse sets of ideas (conceptual, procedural and intuitive) from their knowledge system. &&  Search for Ideas in Associative Memory (SIAM) && \cite{nijstad2002cognitive,george2020need,connolly1993effectiveness,kohn2011collaborative,fink2010enhancing,rennie1996placing}\\

\multirow{3}{2.5cm}{Mechanistic Explanations (ME)} & (ME1) cues students to generate mechanistic explanation(s), then it is more likely to invoke references to everyday and curricular knowledge. && Representational formats of mechanism knowledge && \cite{disessa1993toward,russ2008recognizing,hammer1995student,sirnoorkar2022sensemaking,methodssirnoorkar} \\

 & (ME2) to generate mechanistic explanation(s), then it is more likely to elicit mechanistic accounts. &&&& \\

 & (ME3)  to generate mechanistic explanation(s), then it is more likely to cue generation and connection of diverse ideas from their knowledge system. && Theory of mechanistic reasoning && \cite{russ2008recognizing,de2022students,bachtiar2021stimulating,de2022students,bachtiar2021stimulating} \\ 

 \multirow{2}{2.5cm}{Multiple Representations (MR)} & (MR1) cues students to engage with multiple representations, then it is more likely to cue generation and connection of ideas from their knowledge system.   && Cognitive Theory of Multimedia Learning (CTML) & & \cite{de2012representation,podolefsky2006use,van2001multiple} \\

 & (MR2) cues students to engage with multiple representations, then it is more likely to nudge them in seeking coherence between ideas  && && \cite{cox1999representation,gire2015structural} \\

 \multirow{2}{2.5cm}{Physical Interpretation (PI)} & (PI1) cues students to interpret mathematical expressions in light of physical systems, then it is likely to facilitate generation and connection of ideas.  && 
Mapping account/ Interpretation/ Inferential Conception/ Inferential Function && \cite{sherin2001students,arcavi1994symbol,kuo2013students} \\

& (PI2)  cues students to interpret mathematical expressions in light of physical systems, then it is likely to nudge them in seeking coherence between ideas. && && \cite{kuo2020assessing,dreyfus2017mathematical,ACER}  \\

\end{tabular}
\end{ruledtabular}
\end{center}
\end{table*}

\section{Discussion}
\label{sec:discussion}

\subsection{Operationalizing conjecture mapping in the context of task-design in physics}

We operationalize conjecture mapping -- a framework in design-based research -- in the context of identifying the assessment task features that promote sensemaking in physics ({\bf RQ1}). Based on the literature's description of the sensemaking process, we note the sensemaking elements (the set of activities) that constitute sensemaking. The sensemaking elements correspond to the {\em mediating processes} of our conjecture map - the set of interactions and artifacts produced by participants while engaging with the designed learning environment. Our {\em high level conjecture} takes the form: ``a set of task features which elicit the sensemaking elements increase the likelihood of students sensemaking". These task features then correspond to the {\em embodiment} component of our conjecture map - the material features which elicit the sensemaking elements of sensemaking. The arguments substantiating the embodiment, i.e. our {\em design conjectures}, have been detailed in Sections~\ref{sec:real-world} to~\ref{sec:physical-interpretation}. Lastly, we note the {theoretical conjectures} about the {\em outcomes} of engaging in the sensemaking process from the literature (Section~\ref{subsec:outcomes}). Figure~\ref{fig:conjecture-map-task-features} highlights the contextual operationalization of conjecture mapping to our study.    

\subsection{Identifying task features that increase the likelihood of students sensemaking in physics}

We identify the task features which increase the likelihood of students sensemaking in physics ({\bf RQ2}) by leveraging contemporary theoretical ideas from cognitive psychology, education, and philosophy of science. These features include tasks cuing students about: (i) the presence of real-world context(s), (ii) to unpack the underlying mechanism of a phenomenon, (iii) to engage with multiple representations, and (iv) to physically interpret mathematical expressions. The identified features complement the contemporary pedagogical efforts in supporting students in making sense of their surrounding world using curricular ideas. 

Several studies in science education have examined students' reasoning while engaging with real-world contexts (our first task feature). In addition to developing context-based pedagogical materials~\cite{heller1992teaching,loffler2016modellanwendung,ramalingam2017pisa,price2022accurate}, researchers have analyzed students' cognitive, meta-cognitive, and affective behaviors while engaging with such materials~\cite{pozas2020effects,loffler2018students,park2004analysing,rennie1996placing,taasoobshirazi2008review,park2004analysing,pozas2020effects,bennett2007bringing}. These studies have noted context-based problems to enhance students' situational interest~\cite{rennie1996placing}, motivation~\cite{yager2005exemplary,rayner2005reflections,heller1992teaching,duke2006authentic}, along with improving attitudes towards science learning~\cite{kurbanouglu2015effect}. Our work adds increased chances of engaging in sensemaking to this growing list. Ogilivie~\cite{ogilvie2009changes} indeed notes context rich, open-ended problems to be fertile grounds for students to notice inconsistencies in their knowledge systems - a crucial feature of the sensemaking process. 

Researchers have also explored students' engagement with multiple representations (our third task feature) in physics. Coordination between representations -- referred to as ``representational fluency'' -- has been noted to assist students in invoking conceptual ideas not specified in the problem statement~\cite{gire2015structural,schmidgall2019learners}, leveraging information highlighted by representation(s)~\cite{podolefsky2006use}, and facilitating organization, prioritization, and communication of the contextual information~\cite{susac2019role,vignal2021physics,cox1995supporting,cox1999representation}. Our work adds to this list by noting that engaging with multiple representations leads to generating ideas and establishing coherence between the ideas thereby facilitating sensemaking.

Recent investigations have also explored the close association between sensemaking and modeling. Sirnoorkar {\em et al.}~\cite{sirnoorkar2022sensemaking} note assembling of prior knowledge during sensemaking to entail construction of mental models about the target systems. The authors also note addressing and resolving the perceived inconsistencies during sensemaking to entail coherence seeking in the models, and testing them in light of their target systems. Our identified task features complement these observations by facilitating promotion of sensemaking through modeling. Real-world systems (our first task feature) specify the nature of target systems, which when modeled, can increase the likelihood of students sensemaking in physics. Similarly, coordinating multiple representations, and physically interpreting mathematical results (the last two task features) specify the ways of establishing  coherence and testing the merit of the models.        

\section{Conclusion}
\label{sec:conclusion}

We make a theoretical argument that to promote sensemaking, tasks should cue students to unpack the underlying mechanism of a real-world phenomenon by coordinating multiple representations and by physically interpreting mathematical expressions. We make this argument by leveraging existing theoretical perspectives on the cognitive features of sensemaking, and by adopting conjecture mapping~\cite{sandoval2014conjecture}. 

One of the primary contributions of this work involves adopting an agent-based approach in articulating task-design arguments in physics. Research on task design has traditionally focused on the valued objectives of researchers by overlooking the role of contextual features in influencing students' engagement with tasks. The current work presents an exemplar case by simultaneously attending to the valued objectives of the researchers along with accounting for the students contextual factors. While our theoretical approach on deducing sensemaking elements from its definition reflects the researchers' valued objectives in task design, the vocabulary adopted in making task-related arguments reflect the consideration of students' contextual factors. 

Our work also makes contribution to the contemporary literature by operationalizing conjecture mapping in the context of task design in physics. This technique has been traditionally employed in designing learning environments such as (but not limited to) vocational training~\cite{boelens2020conjecture}, online or hybrid learning~\cite{wozniak2015conjecture,tawfik2020role}, or pedagogy in informal communities~\cite{lee2018conjecture}. The current work leverages this framework in designing physics tasks. Operationalization of this framework also brings together the broad literature on sensemaking (Refer Section~\ref{sec:lit-review}). While the `embodiment' and `mediating processes' (Section~\ref{subsec:conjecture}) encompass the theoretical and analytical views on sensemaking, the `outcomes' embodies the literature on the effects of sensemaking. 

For instructors, this study provides a generalized framework for designing assignment/examination questions, or crafting examples for classroom discussions that can promote sensemaking. The generalized nature of these task features provide avenues for instructors to design tasks based on their local learning objectives and curricula, thereby facilitating their agency~\cite{strubbe2020beyond,miller2018addressing}.

For researchers, the current work describes a methodology for identifying task features, which can be extended to promote other valued epistemic practices such as argumentation or modeling~\cite{k-12framework}. Our proposed methodology -- extracting salient features of a cognitive process from its definition, and back-tracking the task characteristics -- can contribute to the community's efforts in promoting valued epistemic practices in our classrooms. Additionally, there is an increasing traction of investigations on sensemaking in laboratories~\cite{may2022student,sundstrom2020problematizing}. Researchers can also extend our methodology in identifying features of activities or tasks that promote sensemaking during experimentation. 

Since we do not claim the identified features to be exhaustive, researchers can expand on the proposed list, or can further investigate the conditions in which the identified task features are effective. Contemporary reports on our task features do indicate certain accompanying constraints. Heckler~\cite{heckler2010some} notes explicit prompting on constructing representations may cue protocol-based approaches to learning as opposed to intuitive engagement with the content. Similarly, researchers have noted real-world contexts in tasks to elicit subjective judgements about the scenarios~\cite{park2004analysing}, and initiating gender-based disparity in performances~\cite{mccullough2004gender}. 

This study also opens up avenues to explore the interaction of the proposed task features with their structural features. Research on task design has noted activities such as designing experiments, or modeling complex systems to differ from solving the typical end-of-the chapter physics problems. Unlike the former (referred as ill-structured problems), the latter (well-structured) tasks have a well-defined protocol for initiating, proceeding, and terminating the activity~\cite{reed2016structure,shin2003predictors,pulgar2019performance,pulgar2020social,law2020understanding}. The current study paves way for researchers to probe the influence of our identified task features in the context of well- and ill-structured problems. 

However, a number of caveats accompany the claims made in this paper. Firstly, the sensemaking process, according to our adopted definition, is driven by a perceived gap in one's knowledge system~\cite{defining}. This noticing of inconsistencies depends on prior knowledge, awareness, self-evaluation, and approaches students employ while reasoning about the given scenario~\cite{anderson2022grasping}.  Our proposed task features do not attend to this crucial element of the sensemaking process due to its contextual and meta-cognitive nature. Additionally, our objective is neither to argue that the proposed task features {\em necessarily} engage students into sensemaking, nor these features to be the {\em only ones} to promote sensemaking. Rather, we make a modest argument that the identified features which when present in a task {\em together}, enhance the likelihood of students sensemaking in physics.   

Future work would involve analyzing students' responses on the tasks embedding the four proposed task features, and validating (or refining) our conjecture map (Figure~\ref{fig:conjecture-map-task-features}). We also intend to explore the effect of the task features in open-ended tasks as compared to the ones with scaffolds.

\section{Acknowledgements}
We would like to thank Dean Zollman, Bethany Wilcox, Brandi Lohman, and Bill Bridges for their valuable insights. This material is based upon work supported by the National Science Foundation under Grant No. 1726360.

\bibliography{ref}

\end{document}